\documentclass[final,3p,times]{elsarticle}

\usepackage{lineno,hyperref}
\usepackage{amssymb}
\usepackage{amsmath}
\journal{Physica A}

\def\Eq#1{Eq.~(\ref{#1})}
\def\Eqs#1{Eqs.~(\ref{#1})}

\def\no{\nonumber\\}
\def\r{\rangle\!\rangle}
\def\l{\langle\!\langle}
\def\>{\rangle}
\def\<{\langle}
\def\u#1{\underline{#1\!}\,	}

\def\half{\frac{1}{2}}

\def\xt{\tilde{x}}
\def\lam{\lambda}
\def\kap{\kappa}
\def\w{\omega}
\def\sig{\sigma}
\def\Del{\Delta}
\def\del{\delta}
\def\gam{\gamma}

\def\al{\alpha}

\def\th{\theta}

\def\T{\text{T}}
\def\d{\partial}
\def\KL{\text{KL}}
\def\CL{\text{CL}}
\def\HPZ{\text{HPZ}}
\def\Qt{\tilde{Q}}
\def\rt{\tilde{r}}
\def\Qb{\bar{Q}}
\def\rb{\bar{r}}
\def\eq{\text{ss}}

\begin{document}

\begin{frontmatter}

\title{Eigenvalues of the Liouvillians of Quantum Master Equation for a Harmonic Oscillator}

\author[1]{B. A. Tay}
\ead{BuangAnn.Tay@nottingham.edu.my}
\address[1]{Department of Foundation in Engineering, Faculty of Science and Engineering, University of Nottingham Malaysia, Jalan Broga, 43500 Semenyih, Selangor, Malaysia}

\date{\today}

\begin{abstract}
The eigenvalues of the Liouvillians of Markovian master equation for a harmonic oscillator have a generic form. The Liouvillians considered are quadratic in the position coordinates or creation and annihilation operators, as well as having positive renormalized frequencies. We prove this by showing that a generic Liouvillian of this form can be similarly related to the Liouvillian of the Kossakowski--Lindblad equation, whose eigenvalues are already known. The left and right eigenfunctions of the generic Liouvillian also form a complete and biorthogonal set. Examples of similarly related right eigenfunctions are given.
\end{abstract}


\end{frontmatter}


\section{Introduction}

The master equation approach is widely used to study various quantum systems under the influence of an environment it is in contact with \cite{Weiss,Breuer}. It is applied to the studies of quantum transport processes \cite{May11}, quantum information processing \cite{Nielsen}, quantum entanglement \cite{Dodd04Halliwell}, quantum optics \cite{Gardiner,Walls08}, and the more formal aspects of quantum decoherence and quantum Brownian motion \cite{CL83,HPZ92,Zurek03,Decoh03}.

The most commonly used master equation is the Kossakowski-Lindblad (KL) equation that appears in various setups as a result of the Markov approximation \cite{Breuer}.
Various methods have been developed to obtain its solution. It can be solved using phase space methods, for example, in the coherent states representation \cite{Agarwal73} or through the Wigner representation \cite{Tay08}. There are also various operator methods to solve the equation, such as by means of the damping basis \cite{Briegel93}, the eigenprojection method \cite{Honda10}, or by utilizing the internal symmetry of the underlying Liouville space \cite{Tay17b}. There is numerical method that makes use of the Gaussian ansatz \cite{Decoh03}.
The eigenvalues of the Liouvillian of the KL equation, or the generator of time evolution,  were obtained in Refs.~\cite{Briegel93,Tay08,Honda10}.

Other commonly discussed master equations are, for example, the Caldeira-Leggett (CL) equation \cite{CL83} and the Hu-Paz-Zhang (HPZ) equation \cite{HPZ92}. The CL equation was initially derived to study quantum tunnelling effect in the high temperature limit \cite{CL83}. Its solution could be obtained in the position coordinates \cite{Roy99} and through operator method \cite{Chen12}. The CL equation was generalized to study the positivity in the time evolution of master equation \cite{Diosi93a,Diosi93b}. It was also used to study quantum entanglement between two oscillators \cite{Dodd04Halliwell}. On the other hand, the Hu-Paz-Zhang (HPZ) equation was derived to study quantum Brownian motion and decoherence \cite{HPZ92}. Its solution was also obtained and studied \cite{Ford01,Fleming11}.

Generic master equation for a harmonic oscillator was also developed. It could be derived from the Gauss Markov process \cite{Talkner81}. It was used to study the constraint on the diffusion coefficients of master equation \cite{Dekker84}. The solution of generic master equation with time-dependent coefficients was also used to determine the positivity condition on the coefficients of the master equation \cite{Tameshtit13}.

However, as far as we know, the eigenvalues of the Liouvillians other than the KL equation are not known. Solving the KL equation under the damping basis \cite{Briegel93} and the eigenprojection method \cite{Honda10} only requires the use of three basis operators. The number of basis operators involved increases up to seven for generic master equation. It is therefore a challenge to obtain the eigenvalues directly using existing methods.

Recently, we investigated the internal symmetry in the Liouville space of a harmonic oscillator \cite{Tay17} through generalized Bogoliubov transformation \cite{Khanna06b,Tay07}.
Here, we exploit the internal symmetry operator to show that the Liouvillian of the KL equation whose eigenvalues are known is similar to the generic Liouvillian of quadratic nature, for which the Liouvillians of the CL equation and the Markovian limit of the HPZ equation are special cases. They therefore share the same eigenvalues. In this way, we bypass the complications of solving for the eigenvalues directly through existing operator methods.

Since the master equations are developed to describe different physical situations, we expect them to exhibit different types of dissipative behaviors. Therefore, the result that their eigenvalues turn up to have the same generic form which is independent of the diffusion constants of the master equation is quite unexpected. The eigenvalue depends only on the renormalized frequency of the harmonic oscillator and the relaxation constant of the system.i

The organization of this paper is as follows. In Sec.~\ref{SecKL}, we first discuss the known results in the eigenvalue problem of the Liouvillian of the KL equation to lay down the setting of our subsequent dicsussions. This is followed by introducing the generic Liouvillian of the master equation for a harmonic oscillator and the internal symmetry operator in the Liouville space in Sec.~\ref{SecSimTransf}. In Sec.~\ref{SecProof}, using the symmetry operator we prove that the generic Liouvillian is similar to the Liouvillian of the KL equation. We obtain the closed form of the right eigenfunctions in Sec.~\ref{SecRightFunc} before we conclude our investigations in Sec.~\ref{SecConclusion}. Examples of the right eigenfunctions are left to the appendices.

\section{Eigenvalues and eigenfunctions of the Liouvillian of Kossakowski-Lindblad (KL) equation}
\label{SecKL}

The Kossakowski-Lindblad (KL) equation for a harmonic oscillator $\d \rho/\d t=-K\rho$ has the Liouvillian
\begin{align}   \label{KKL}
    K_\KL&=2\w_o iL_0+\gam (O_0-I/2)-2\gam b O_+\,,
\end{align}
where $\w_0$ is the natural frequency of the oscillator, $\gam$ is the relaxation constant of the system, and $b=\frac{1}{2}\coth[\hbar\w_0/(2kT)]$ is a parameter of the temperature $T$ of the bath.
In \Eq{KKL}, $iL_0, O_0-I/2$ and $O_+$ are three basis operators out of a set of seven that preserve the trace and hermiticity of density matrix \cite{Tay17,Tay17b}.

In the center and relative coordinates defined by
\begin{align} \label{Qrxx}
    Q&\equiv\frac{1}{2}(x+\xt)\,,   \qquad r\equiv x-\xt\,,
\end{align}
respectively, which are related to the dimensionless position coordinates of the bra- $|\xt\>$ and the ket-space $\<x|$, the operators take the following form \cite{Tay17b}
\begin{align}   \label{iL0}
    iL_0 = \frac{i}{2}\left(-\frac{\d^2}{\d Q\d r}+Qr\right)\,,\qquad
    O_0-I/2 = -\frac{1}{2}\left(\frac{\d}{\d
    Q}Q-r\frac{\d}{\d r}\right)\,,\qquad
    O_+ = \frac{1}{4}\left(\frac{\d^2}{\d Q^2}-r^2\right)\,.
\end{align}
The right eigenvalue problem
\begin{align}   \label{Kfmn}
    K_\KL|f^\pm_{mn}\r&=\lam^\pm_{mn}|f^\pm_{mn}\r
\end{align}
was solved in Ref.~\cite{Tay08} with the eigenvalues
\begin{align}   \label{lam}
    \lam^\pm_{mn}&=\pm i n\w_0+(m-n/2)\gam\,,   \qquad 0\leq n\leq m\,.
\end{align}
The eigenvalues were also obtained earlier in Ref.~\cite{Briegel93} and later in Ref.~\cite{Honda10}.
The right eigenvectors $|f^\pm_{mn}\r$ together with their left counterparts $\l g^\pm_{mn}|$ form a complete and biorthonormal set \cite{Tay08},
\begin{align}   \label{fg}
    \sum_{m=0}^\infty \sum_{n=0}^m \sum_{\sig=\pm}|f^\sig_{mn}\r\l g^\sig_{mn}|=I\,,\\
    \l g^\sig_{mn}|f^{\sig'}_{m'n'}\r =\del_{m,m'}\del_{n,n'}\del_{\sig,\sig'}\,.\label{fgdel}
\end{align}
In the position coordinates, the right eigenfunctions are
\begin{align}   \label{fQr}
    f^\pm_{mn}(Q,r)
                =f_{00}(Q,r) \sum_{\mu=0}^{m-n}\sum_{\nu=0}^{\mu}
                    \sum_{\sig=0}^{n} c^{\pm\mu\nu\sig}_{mn} \left(\frac{Q}{ \sqrt{2b}} \right)^{2(\mu-\nu)+n-\sig}
                     H_{2\nu+\sig}\left( \sqrt{ \frac{b}{2}} \,  r \right)\,,
\end{align}
with the stationary state
\begin{align}  \label{f00}
            f_{00}(Q,r)=\frac{1}{\sqrt{2 \pi b}}
           \,  e^{-Q^2/(2b)-br^2/2} \,.
\end{align}
In \Eq{fQr}, $H_{2\nu+\sig}$ denotes the Hermite polynomials, and
\begin{align}  \label{cmnpm}
            c^{\pm\mu\nu\sig}_{mn} &= (\pm 1)^{n+\sig}   \frac{ (-1)^{\mu+\nu} }{ i^n 2^{2\nu+\sig} \mu !} \sqrt{\frac{(m-n)!}{m!} }  \left(\begin{matrix} m \cr n+\mu \end{matrix} \right)
               \left(\begin{matrix} \mu \cr \nu \end{matrix} \right)
               \left(\begin{matrix} n \cr \sig \end{matrix} \right)
\end{align}
are coefficients.
The left eigenfunctions are generalized functions \cite{Bohm78,Tay08}. Since they are not the main focus of this work, we refer the readers to Ref.~\cite{Tay08} for the details.

Under a similarity transformation by an operator $S$, the Liouvillian of the KL equation is transformed to $K'\equiv SK_\KL S^{-1}$. The eigenvalue problem then becomes
\begin{align}   \label{S}
   K'|{f'}^\pm_{mn}\r&=\lam^\pm_{mn}|{f'}^\pm_{mn}\r\,,
\end{align}
with the transformed right eigenvector $|{f'}^\pm_{mn}\r\equiv S|f^\pm_{mn}\r$, whereas the eigenvalue remains the same.
The left eigenvector transforms into $\l {g'}^\sig_{mn}|\equiv\l g^\sig_{mn}|S^{-1}$. As a result, the completeness and biorthonormal relations in \Eqs{fg}-\eqref{fgdel} are still valid under the similarity transformation.

In the following, we will show that the operator $S$ can indeed be constructed. We will show that the Liouvillian of generic master equation that are quadratic in the position coordinates, or in the creation and annihilation operators, can be related to $K_\KL$ by a similarity transformation. This includes the familiar Caldeira-Leggett (CL) equation \cite{CL83} and the Hu-Paz-Zhang (HPZ) equation \cite{HPZ92} in the Markovian limit. Hence, the eigenvalues of the class of quadratic master equation possess the same generic form.

\section{Similarity transformation}
\label{SecSimTransf}

\subsection{Generic Liouvillian of quiantum master equation}
\label{SecGenL}

The Liouvillian of quadratic master equation for a harmonic oscillator that preserves the trace and hermiticity of density matrix has the generic form \cite{Tay17,Talkner81,Tameshtit13} (in our notation)
\begin{align}   \label{GenK}
    K&=K_0+\gam (O_0-I/2)+K_1\,,
\end{align}
where we divide $K$ into a unitary part
\begin{align}   \label{K0}
    K_0&=h_0iL_0+h_1iM_1+h_2iM_2\,,
\end{align}
and a dissipative part. In anticipation of the transformation properties of the operators, we further separate the dissipative part into two components, namely, $O_0-I/2$ and
\begin{align}   \label{Kd}
    K_1&=g_+O_++g_1L_{1+}+g_2L_{2+}\,.
\end{align}
The $h_i$s and $g_i$s are real coefficients of the master equation.
Besides the three operators listed in \Eq{iL0}, the other four operators are \cite{Tay17,Tay17b}
\begin{align}   \label{M1}
    iM_1= \frac{i}{2}\left(\frac{\d^2}{\d Q\d r}+Qr\right)\,,\qquad
    iM_2= -\frac{1}{2}\left(\frac{\d}{\d Q}Q+r\frac{\d}{\d r}\right)\,,\qquad
    L_{1+}= -\frac{1}{4}\left(\frac{\d^2}{\d Q^2}+r^2\right)\,, \qquad
    L_{2+}= -\frac{i}{2}r\frac{\d}{\d Q}\,.
\end{align}

To form a complete set of quadratic operators in the Liouville space of a harmonic oscillator, we need to supplement the seven operators listed above by three more operators, $O_-, L_{1-}$ and $L_{2-}$, cf.~Ref.~\cite{Tay17} for the details. However, the latter operators do not preserve the trace of density matrix, though they preserve its hermiticity. Consequently, they do not take part in the dynamics and are absent from the Liouvillian.

The seven operators in \Eqs{iL0} and \eqref{M1} are closed under the commutator brackets \cite{Tay17}. The commutation relations between the operators can be found in Table 1 of Ref.~\cite{Tay17} or in Table B.3 of Ref.~\cite{Tay19b}.
Treating the seven operators as generators of transformation, we exponentiate them over real parameters to produce general transformation in the Liouville space of a harmonic oscillator. Three of them,
\begin{align}   \label{U0}
    U_0(\th)\equiv e^{\th iL_0}\,, \qquad
    U_1(\phi)\equiv e^{\phi iM_1}\,, \qquad
    U_2(\psi)\equiv e^{\psi iM_2}\,,
\end{align}
are ordinary transformation that we are familiar with, i.e., they are unitary. $U_0(\th)$ is reminiscent of the time evolution operator of a free oscillator, whereas $U_1(\phi)$ and $U_2(\psi)$ are the squeezed operators \cite{Gardiner,Walls83}.

The rest of the four operators are
\begin{align}   \label{G0}
    G_0(\al)\equiv e^{\al (O_0-I/2)}\,, \qquad G_+(\eta_0)\equiv e^{\eta_0 O_+}\,, \qquad G_1(\eta_1)\equiv e^{\eta_1 L_{1+}}\,, \qquad G_2(\eta_2)\equiv e^{\eta_2 L_{2+}}\,.
\end{align}
They generate nonunitary transformation \cite{Tay17}. For this reason, we call them general tranaformation to distinguished them from the ordinary ones. Nonetheless, all of the $U_i$s and $G_i$s are invertible. Their inverse are $U_0^{-1}(\th)=U_0(-\th), G_0^{-1}(\al)=G_0(-\al),$ and etc.

Using the commutation relations of the operators, we can work out the effects of similarity transformation induced by each of the operators on the generic Liouvillian \eqref{GenK}-\eqref{Kd}. We reproduce the results of Ref.~\cite{Tay17} below for the convenience of our discussion. They are
\begin{subequations}
\begin{align}   \label{URK}
    e^{\th iL_0} K e^{-\th iL_0}&= h_0iL_0+ (h_1 \cos\th + h_2 \sin\th)iM_1+ (h_2\cos\th -h_1 \sin\th)iM_2+ \gam(O_0-I/2)\no
    &\quad+ g_+O_++ (g_1 \cos\th + g_2 \sin\th)L_{1+}+( g_2 \cos\th - g_1 \sin\th)L_{2+}\,,
\\  \label{U1K}
    e^{\phi iM_1} K  e^{-\phi iM_1}
    &=(h _0\cosh\phi+h _2\sinh\phi)iL_0+ h _1iM_1+ (h _2\cosh\phi +h _0 \sinh\phi)iM_2 +\gam(O_0-I/2)\no
    &\quad+( g_+\cosh\phi+g_2\sinh\phi)O_++ g_1L_{1+}+( g_2 \cosh\phi + g_+ \sinh\phi)L_{2+}\,,
\\   \label{U2K}
     e^{\psi iM_2} K  e^{-\psi iM_2}
    &=(h _0\cosh\psi-h _1\sinh\psi)iL_0+(h _1\cosh\psi -h _0 \sinh\psi)iM_1+ h _2iM_2  +\gam(O_0-I/2)\no
    &\quad+ (g_+\cosh\psi-g_1\sinh\psi)O_+ +(g_1\cosh\psi - g_+ \sinh\psi)L_{1+}+ g_2L_{2+}\,,
\\  \label{U0K}
    e^{\al (O_0-I/2)} K e^{-\al (O_0-I/2)}&=h _0iL_0+ h _1iM_1+ h _2 iM_2  +\gam(O_0-I/2)\no
    &\quad + e^\al g_+O_+  + e^\al g_1L_{1+}+ e^\al g_2 L_{2+}\,,
\\  \label{U+K}
    e^{\eta_0 O_+} K e^{-\eta_0 O_+}&=h _0iL_0+ h _1iM_1+ h _2iM_2+\gam(O_0-I/2)\no
    &\quad + (g_+ -\eta_0 \gam)O_+ +(g_1+\eta_0 h _2)L_{1+}+( g_2-\eta_0 h _1)L_{2+}\,,
\\  \label{U1+K}
   e^{\eta_1 L_{1+}} K e^{-\eta_1 L_{1+}}&=h _0iL_0+ h _1iM_1+ h _2iM_2 +\gam (O_0-I/2)\no
    &\quad + (g_+ +\eta_1 h _2)O_+  +(g_1- \eta_1 \gam)L_{1+}+( g_2+\eta_1 h _0)L_{2+}\,,
\\  \label{U2+K}
    e^{\eta_2 L_{2+}} K e^{-\eta_2 L_{2+}}&=h _0iL_0+ h _1iM_1+ h _2iM_2 +\gam(O_0-I/2)\no
    &\quad + (g_+ -\eta_2 h _1)O_+ +(g_1- \eta_2 h _0)L_{1+}+( g_2-\eta_2 \gam)L_{2+}\,.
\end{align}
\end{subequations}

A few important observations follows. First, the coefficient of $O_0-I/2$ is invariant under the similarity transformation. This coefficient is the relaxation constant of the system. Second, similarity transformation induced by $G_i$ does not affect the unitary part $K_0$ of the Liouvillian. Armed with these observations, we will show that the generic Liouvillian \eqref{GenK} can be similarly related to $K_\KL$ in two steps, i.e., $S_2S_1KS^{-1}_1S^{-1}_2=K_\KL$, where $S_1$ and $S_2$ are defined below.
\begin{enumerate}[{Step 1.}]
\item We first show that a similarity transformation induced by $S_1\equiv U_1(\phi)U_0(\th)$ (or the products of any two copies of $U_i$) can bring the unitary part $K_0$ of the generic Liouvillian \eqref{GenK}-\eqref{Kd} into the unitary part of $K_\KL$ \eqref{KKL}, which has $h_0=2\w_0$ and $h_1=0=h_2$.

\item We then show that with $S_2\equiv U_2(\eta_2)U_1(\eta_1)U_+(\eta_0)$ (where $O_+, L_{1+}$ and $L_{2+}$ mutually commute \cite{Tay17}), the coefficients $g_i$ of $K_1$ can be brought by the similarity transformation into arbitrary $g'_i$ of any $K'_d$, for which the dissipative part of $K_\KL$ is a special case with $g_+=-2\gam b$, and $g_1=0=g_2$.
\end{enumerate}
If necessary, applying one more similarity transformation under $G_0(\al)$ can further scale $g'_+$ to any desired number $g'_+ e^{\al}$ with the same sign, cf.~\Eq{U0K}. In this way we show that the Liouvillian of generic master equation can be related by a series of similarity transformation to $K_\KL$. Hence, the generic Liouvillian possesses the same set of eigenvalues $\lam^\pm_{mn}$ \eqref{lam} as $K_\KL$.

\section{Proof}
\label{SecProof}

\subsection{Step 1 by ordinary transformation}
\label{SecStep1}

The effects of the ordinary transformation $U_i$ on the Liouvillian can be summarised compactly by means of the following matrices,
\begin{align}   \label{eth}
    \underline{U}_0(\th)&=\left(\begin{array}{ccc}
                    1 & 0 & 0 \\
                    0 & \cos\th & \sin\th \\
                    0 & -\sin\th & \cos\th
                \end{array}\right)\,,\qquad
    \underline{U}_1(\phi)&=\left(\begin{array}{ccc}
                    \cosh\phi & 0 & \sinh\phi \\
                    0 & 1 & 0 \\
                    \sinh\phi & 0 & \cosh\phi
                \end{array}\right)\,,\qquad
    \underline{U}_2(\psi)&=\left(\begin{array}{ccc}
                    \cosh\psi & -\sinh\psi & 0 \\
                    -\sinh\psi & \cosh\psi & 0 \\
                    0 & 0 & 1
                \end{array}\right)\,.
\end{align}
The matrices act on the column vectors $\vec{h}=(h_0,h_1,h_2)^\T$ and $\vec{g}=(g_+,g_1,g_2)^\T$ by matrix multiplication to yield the corresponding coefficients of the transformed Liouvillian. We use $\T$ to denote matrix transpose.

Let us first focus on $\vec{h}$. Given any $\vec{h}$ of a Liouvillian, we can bring it into $\vec{h}_\KL=(2\w_0,0,0)^\T$ of the KL equation via $\vec{h}_\KL=\underline{U}^{-1}_1(\phi)\cdot \underline{U}^{-1}_0(\th)\cdot\vec{h}$, provided the parameters of the transformation satisfy
\begin{align}   \label{S1thphi}
    \tan\th=\frac{h_1}{h_2}\,,\qquad
    \tanh\phi=\sqrt{\left(\frac{h_1}{h_0}\right)^2
            +\left(\frac{h_2}{h_0}\right)^2}\,.
\end{align}
The equations show that $\th$ and $\phi$ are always real. Hence, the transformation exists for any $\vec{h}$. Furthermore, ordinary transformation preserves the length of $\vec{h}$ with the metric $(1,-1,-1)$, cf.~ Ref.~\cite{Tay17,Tay17b}. Hence, we have $\vec{h}^2_\KL=\vec{h}^2$, or
\begin{align}   \label{wh012}
    4\w^2_0=h^2_0-h^2_1-h^2_2\,.
\end{align}
The reality of the natural frequency $\w_0$ then requires
\begin{align}   \label{h0>h12}
        h^2_0-h^2_1-h^2_2\geq 0
\end{align}
in the Liouvillian.

From the solution of generic master equation in Ref.~\cite{Tay17b}, we know that by parameterizing $h_0$ as $h_0=2\w_0'$, the coefficient $h_1$ of $iM_1$ renormalizes the natural frequency of the harmonic oscillator to $\w^{\prime 2}_0-h^2_1/4$. The coefficient $h_2$ of $iM_2$ then determines \cite{Tay17b} whether the oscillator is underdamped $\w^2_0=\w^{\prime 2}_0-h^2_1/4-h^2_2/4>0$, critically damped $\w^2_0=0$, or overdamped $\w^2_0<0$.

In the overdamped case, the master equation may not have stationary solution, see the analysis in Sec.~6.1 of Ref.~\cite{Tay17b}. Another example where the master equation does not have a thermal equilibrium solution was provided in Sec.~6 of Ref.~\cite{Tay10}, which considered a particle diffusing through a heat bath in a non-oscillatory motion. It has plane-wave like solution in the position coordinate and it has continuous eigenvalues. In this situation, we have the critically damped condition $\w_0=0$, with $h_0=2\w_0'$ and $h_1=-2\w_0'$. We avoid such situations by imposing the requirement \eqref{h0>h12}, taking note that, however, our results are still applicable to the critically damped situation $\w_0=0$ in which $h_0=h_1=h_2=0$. This is reminiscent of the interaction picture, where we go into the rotating frame of the harmonic oscillator.

The similarity transformation brings the vector $\vec{g}$ into
\begin{align}   \label{Ug}
    \vec{g}_U=\u{U}^{-1}_1(\phi)\cdot \u{U}^{-1}_0(\th) \cdot \vec{g}\,.
\end{align}

\subsection{Step 2 by general transformation}
\label{SecStep2}

We observe from \Eqs{U+K}-\eqref{U2+K} that under similarity transformation induced by $G_+(\eta_0), G_1(\eta_1)$ and $G_2(\eta_2)$, the vector $\vec{g}_U$ is translated into
\begin{align}   \label{Geta}
    \vec{g}'&=\vec{g}_U + \u{G} \cdot \vec{\eta}\,,\\
    \u{G} &=\left(\begin{array}{ccc}
                    -\gam & h_2 & -h_1 \\
                    h_2 & -\gam & -h_0 \\
                    -h_1 & h_0 & -\gam \\
                \end{array}\right)\,,
\end{align}
where $\vec{\eta}=(\eta_0,\eta_1,\eta_2)^\T$ are parameters of the general transformation. The determinant of $\u{G}$ is
\begin{align}   \label{DetD}
    \text{det } \u{G}=-\gam(h^2_0-h^2_1-h^2_2+\gam^2)\,.
\end{align}
The existence of the inverse of $\u{G}$ requires $h^2_0-h^2_1-h^2_2+ \gam^2=4\w_0^2+\gam^2\neq 0$, which is always fulfilled by virtue of requirement \eqref{h0>h12} in Step 1.

Then, denoting $\del\vec{g}\equiv\vec{g}'-\vec{g}_U$, we invert $\u{G}$ to solve for the parameters of the general transformation in terms of the known coefficients of the Liouvillian,
\begin{align}   \label{invD}
    \vec{\eta}=\u{G}^{-1}\cdot \del\vec{g}\,.
\end{align}

After Step 1, we obtain a Liouvillian with the $K_0$ part containing only the $iL_0$ component, $\vec{h}_\KL=(2\w_0,0,0)^\T$, where $\w_0$ is given by \Eq{wh012}. Step 2 is completed by setting $\vec{g}'=\vec{g}_\KL=(-2\gam b,0,0)^\T$, to finally transformed $K$ into $K_\KL$ under a series of similarity transformation.
We can also scale the value of $b$ to $be^\al$ by a further similarity transformation under $G_0(\al)$, cf.~\Eq{U0K}.

In summary, we have shown that the generic Liouvillian with coefficients satisfying the inequality \eqref{h0>h12} are similarly related to $K_\KL$. Hence, their eigenvalues are $\lam^\pm_{mn}$ in \Eq{lam}, where $\w_0$ is related to the coefficients $h_0, h_1$ and $h_2$ of the generic Liouvillian \eqref{GenK}-\eqref{Kd} by \Eq{wh012}.
The eigenvalues of the Caldeira--Leggett equation obtained in Ref.~\cite{Tay10} conform with this result. However, we note that the eigenvalues in Ref.~\cite{Tay10} are labelled by different indices and are arranged in  different orders.i

Notice that the rest of the other diffusion coefficients $g_i$ play no role in the eigenvalue.
The result implies that the eigenfunctions $f^\pm_{mn}$ should decay exponentially with the factor $e^{-(m-n/2)\gam t}$, which is independent of the other parameters in the Liouvillian, such as $b$ in \Eq{KKL} that is a function of the temperature of the bath. The result is counter-intuitive with our common sense that a wave packet in contact with a bath of greater temperature should decohere faster.
However, a more detail analysis \cite{Tay08} showed that after summing up all the decaying modes, an estimate of the decoherence time of a wave packet indeed depends on the temperature of the bath.

\section{Right eigenfunctions}
\label{SecRightFunc}

The right eigenvector of the similarly related Liouvillian is formerly given by $|f'^\pm_{mn}\r=S|f^\pm_{mn}\r$, where $|f^\pm_{mn}\r$ is the eigenvector of $K_\KL$ given by \Eqs{fQr}-\eqref{f00} in the position coordinates.
Examples of the first few eigenfunctions of the Cialdeira-Leggett equation were obtained in Ref.~\cite{Tay10} in the Wigner representation by brute force. They exhibit the same factorized structure of a polynomial multiplying the stationary state, cf.~$f^\pm_{mn}$ in \Eq{fQr}.
Owing to this structure, we will see that all of the similarly related right eigenfunctions retain the same form.

We facilitate the subsequent discussions by redefining the position coordinates to
\begin{align}   \label{Qtrt}
    \Qt &\equiv \frac{Q}{\sqrt{2b}}\,, \qquad \rt\equiv \sqrt{\frac{b}{2}} r\,.
\end{align}
The right eigenfunctions of the KL equation can then be cast into a simpler form,
\begin{subequations}
\begin{align}   \label{fmnt}
    f^\pm_{mn}(\Qt,\rt)&= \tilde{\Pi}^\pm_{mn}(\Qt,\rt) f_\KL(\Qt,\rt)\,,\\
    f_\KL(\Qt,\rt)&\equiv\frac{1}{\sqrt{2\pi b}}e^{-\Qt^2-\rt^2}\,,\\
    \tilde{\Pi}^\pm_{mn}(\Qt,\rt)&\equiv \sum_{\mu=0}^{m-n}\sum_{\nu=0}^{\mu}
                    \sum_{\sig=0}^{n} c^{\pm\mu\nu\sig}_{mn} \Qt^{2(\mu-\nu)+n-\sig}
                     H_{2\nu+\sig}(\rt)\,, \label{Pimn}
\end{align}
\end{subequations}
where $f_\KL$ is the stationary state of the KL equation and $\tilde{\Pi}^\pm_{mn}$ denotes the polynomial components of the right eigenfunctions.

A similarity transformation $S$ on the right eigenfunction of the KL equation $f^\pm_{mn}$ then gives the right eigenfunctions of the similarly related Liouvillian,i
\begin{align}   \label{Sf}
    f^{\prime\pm}_{mn}(Q',r')&=Sf^\pm_{mn}(\Qt,\rt)\no
    &= S\tilde{\Pi}^\pm_{mn}(\Qt,\rt)S^{-1}\cdot Sf_\KL(\Qt,\rt)\no  &=\tilde{\Pi}^\pm_{mn}(\hat{Q}',\hat{r}')f_\eq(Q',r')\,,
\end{align}
where $f_\eq$ denotes the stationary state of the similarly related Liouvillian in the normalized coordinates $(Q', r')$, see \ref{AppGauss} and \ref{AppStationary}, whereas
\begin{align}   \label{SqSr}
    \hat{Q}'\equiv S\Qt S^{-1}\,,\qquad  \hat{r}'\equiv S\rt S^{-1}
\end{align}
are differential operators in the new coordinates $(Q', r')$. It is shown in \ref{Appright} that the new coordinates are defined with respect to the coefficients of the $Q^2$ and $r^2$ coordinates in the exponent of the stationary state $f_\eq$. The order of $\hat{Q}'$ and $\hat{r}'$ in $\tilde{\Pi}^\pm_{mn}$ is not important because they commute, $[\hat{Q}',\hat{r}']=S[\Qt,\rt]S^{-1}=0$.
We obtain the explicit expression of the similarly related right eigenfunction $f^{\prime\pm}_{mn}(Q',r')$ by acting the differential operator $\tilde{\Pi}^\pm_{mn}(\hat{Q}',\hat{r}')$ on the stationary state $f_\eq(Q',r')$, cf.~\Eq{Sf}.

Since the stationary state $f_\KL$ is a Gaussian state, similarly related stationary states are also Gaussian states. They can be worked out using the method developed in Sec.~5.2 of Ref.~\cite{Tay17b}.
We report the results in \ref{AppGauss}. Specific examples of the stationary states of the CL equation and the Markovian limit of the HPZ equation are given in \ref{AppStationary}.
To illustrate the method, we also obtain the first few right eigenfunctions of the CL and the HPZ equation in \ref{Appright}.

\section{Conclusion}
\label{SecConclusion}

We have shown that the generic Liouvillian of Markovian master equation for a quantum oscillator with positive renormalized frequency is similar to the Liouvillian of the KL equation, and hence their eigenvalues share the same generic form.
Despite the fact that the harmonic oscillator may couple to the environment in different ways, the eigenvalues depend solely on the renormalized frequency of the harmonic oscillator and the relaxation rate of the system.
The proof is based on the unique structure of the internal symmetry operator of the Liouville space of a harmonic oscillator, which is not likely to repeat itself in other systems, such as finite-level systems.


\appendix

\section{Gaussian states under general transformation}
\label{AppGauss}

Gaussian states have the generic form,
\begin{align}   \label{Gauss}
    f(Q,r)&=\sqrt{\frac{2\mu}{\pi}}e^{-4\mu Q^2/2-\kap iQr-(\mu+\nu)r^2/2}\,,
\end{align}
where $\mu, \kap$ and $\nu$ are real parameters. They are normalized according to
\begin{align}   \label{normGauss}
    \int^\infty_{-\infty} f(x,0) dx =1\,.
\end{align}
The positive semidefiniteness of Gaussian states is determined by the necessary and sufficient conditions, $\mu> 0$ and $\nu\geq 0$ \cite{Simon88,Tay17b}.

Using the method introduced in Sec.~5.2 of Ref.~\cite{Tay17b}, in this appendix we summarized the parameters of the transformed Gaussian states under the transformation $S$,
\begin{align}   \label{f'Qr}
    f'(Q,r)=S\!\!f(Q,r)=\sqrt{\frac{2\mu'}{\pi}}e^{-4\mu' Q^2/2-\kap' iQr-(\mu'+\nu')r^2/2}\,.
\end{align}
The range of the parameters which guarantees the positive semidefiniteness of the Gaussian states will also be given. In the following equations, $\Del$ denotes
\begin{align}   \label{Del}
    \Del\equiv 4\mu(\mu+\nu)+\kap^2\,.
\end{align}

\begin{enumerate}[{(1)}]
\item $S=U_0(\th)$.
\begin{subequations}
\begin{align}   \label{Rprimer}
    \mu'&=\mu/D\,,\\
    \mu'+\nu'&=(\mu+\nu)/D\,,\\
    \kap'&=\kap\cos\th-\half(1-\Del)\sin\th\,,\\
    D&=\big[\cos(\th/2)+\kap\sin(\th/2)\big]^2+4\mu(\mu+\nu)\sin^2(\th/2)\,,
\end{align}
\end{subequations}
$f'(Q,r)$ is positive for all $\th$.

\item $S=U_1(\phi)$.
\begin{subequations}
\begin{align}   \label{Rprime1}
    \mu'&=\mu/D\,,\\
    \mu'+\nu'&=(\mu+\nu)/D\,,\\
    \kap'&=\kap\cosh\phi-\half(1+\Del)\sinh\phi\,,\\
    D&=[\cosh(\phi/2)-\kap\sinh(\phi/2)]^2  +4\mu(\mu+\nu)\sinh^2(\phi/2) \,. \label{detD1}
\end{align}
\end{subequations}
$f'(Q,r)$ is positive for all $\phi$.

\item $S=U_2(\psi)$.
\begin{subequations}
\begin{align}   \label{Rprime2}
    \mu'&=\mu e^{-\psi}\,,\\
    \mu'+\nu'&=(\mu+\nu) e^{-\psi}\,,\\
    \kap'&=\kap e^{-\psi}\,,
\end{align}
\end{subequations}
$f'(Q,r)$ is positive for all $\psi$.

\item $S=G_0(\al)$.
\begin{subequations}
\begin{align}   \label{Rprime0}
    \mu'&=\mu e^{-\al}\,,\\
    \mu'+\nu'&=(\mu+\nu) e^{\al}\,,\\
    \kap'&=\kap \,.\label{Rprime0c}
\end{align}
$f'(Q,r)$ is positive provided
\begin{align}   \label{phi0}
    \half \ln \frac{\mu}{\mu+\nu}\leq \al\,.
\end{align}
\end{subequations}

\item $S=G_+(\eta_0)$.
\begin{subequations}
\begin{align}   \label{Rprime+}
    \mu'&=\mu /D\,,\\
    \mu'+\nu'&=\frac{1}{D}\left[ \mu+\nu+\half(1+\Del)\al+\mu \al^2\right]\,,\\
    \kap'&=\kap/D\,, \label{kapprime+}\\
    D   &=1+2\mu\al\,.\label{detD+}
\end{align}
$f'(Q,r)$ is positive provided
\begin{align}   \label{al+}
    -\frac{1}{2\mu}&<\al\,,\qquad
    \text{and} \qquad \al_+\leq \al \quad \text{or} \quad \al\leq \al_-\,,
\end{align}
in which
\begin{align}   \label{alpm+}
    \al_\pm \equiv \frac{1}{4\mu}\left[-(1+\Del)\pm\sqrt{(1+\Del)^2-16\mu\nu}\right]\,.
\end{align}
\end{subequations}

\item $S=G_1(\eta_1)$.
\begin{subequations}
\begin{align}   \label{Rprime1+}
    \mu'&=\mu /D\,,\\
    \mu'+\nu'&=\frac{1}{D}\left[\mu+\nu +\half(1-\Del)\eta_1-\mu \eta_1^2\right]\,,\\
    \kap'&=\kap/D\,,\label{Rprime1+kap}\\
     D&=1-2\mu\eta_1\,.\label{detD1+}
\end{align}
$f'(Q,r)$ is positive provided
\begin{align}   \label{al1+}
    \eta_1&<\frac{1}{2\mu}\,,\qquad
    \text{and} \qquad \eta_-\leq \eta_1 \leq \eta_+\,,
\end{align}
where
\begin{align}   \label{alpm1+}
    \eta_\pm \equiv \frac{1}{4\mu}\left[1-\Del\pm\sqrt{(1-\Del)^2+16\mu\nu}\right]\,.
\end{align}
\end{subequations}

\item $S=G_2(\eta_2)$.	
\begin{subequations}
\begin{align}   \label{Rprime2+}
    \mu'&=\mu\,,\\
    \mu'+\nu'&=\mu+\nu+\kap\eta_2 -\mu \eta_2^2\,,\\
    \kap'&=\kap-2\mu\eta_2\,.
\end{align}
$f'(Q,r)$ is positive provided
\begin{align}
    \eta_-\leq \eta_2\leq \eta_+\,,\label{alpos2+}
\end{align}
where
\begin{align}
    \eta_\pm &\equiv \frac{1}{2\mu}\left(\kap\pm\sqrt{\kap^2+4\mu\nu}\right)\,.\label{alpm2+}
\end{align}
\end{subequations}

\end{enumerate}

\section{Examples of stationary states}
\label{AppStationary}

Let us use the results of \ref{AppGauss} to work out the stationary states of the Caldeira-Leggett (CL) equation \cite{CL83} and the Hu-Paz-Zhang (HPZ) equation \cite{HPZ92} in the Markovian limit.

\subsection{Caldeira-Leggett (CL) equation}
\label{AppCL}

The Liouvillian of the CL equation is
\begin{align}   \label{KCL}
    K_\CL(\w'_0,b_\CL)&\equiv 2\w'_0 iL_0+\gam(O_0-I/2-iM_2)-2\gam b_\CL(O_++L_{1+})\,.
\end{align}
It can be reached from the KL equation through \cite{Tay17}
\begin{align}
    K_\CL(\w_0',b_\CL)=e^{\phi iM_1}e^{\eta L_{2+}} K_\KL(\w_0,b)e^{-\eta L_{2+}}e^{-\phi iM_1}\,,
\end{align}
provided the parameters satisfy ${\w'}^2_0=\w_0^2+\gam^2/4$, $\sinh\phi=-\gam/(2\w_0)$, $\cosh\phi = \w_0'/\w_0=b/b_\CL$ and $\eta=-2b\tanh\phi$.

Starting with $\mu=1/(4b), \mu+\nu=b$, and $\kap=0$ for $f_{00}(Q,r)$ \eqref{f00}, we subsequently apply the results of item (7) for $G_2(\eta)f_{00}(Q,r)$ and item (2) for $U_1(\phi)f_{00}(Q,r)$ from \ref{AppGauss} to obtain $\mu', \mu'+\nu'$ and $\kap'$. It yields the following results.
\begin{enumerate}[{(A)}]
\item $G_2(\eta)f_{00}(Q,r)$.
\begin{subequations}
\begin{align}   \label{CLA}
    \mu'&=\frac{1}{4b}\,,\\
    \mu'+\nu'&=b-\frac{\eta^2}{4b}\,,\\
    \kap'&=-\frac{\eta}{2b}\,,
\end{align}
with the positive condition
\begin{align}   \label{etapos}
    -\sqrt{\frac{\nu}{\mu}}\leq \eta \leq \sqrt{\frac{\nu}{\mu}}\,.
\end{align}
\end{subequations}
\item $U_1(\phi)G_2(\eta)f_{00}(Q,r)$. Inserting the results of (A) into item (2) of \ref{AppGauss} yields
\begin{subequations}
\begin{align}   \label{CLB}
        \Del&=1\,,\\
        D&=1/\cosh\phi\,,\\
    \mu''&=\frac{1}{4b_\CL}\,,\\
    \mu''+\nu''&=b_\CL\,,\\
    \kap''&=0\,,\\
    b_\CL&\equiv \frac{b}{\cosh\phi}\,.
\end{align}
\end{subequations}
The resulted Gaussian state is positive for all values of $\phi$.
\end{enumerate}
In summary, the stationary state of $K_\CL(\w'_0,b_\CL)$ \eqref{KCL} is also given by \Eq{f00}, but by replacing $b=b_\CL$.

\subsection{Hu-Paz-Zhang (HPZ) equation in the Markovian limit}
\label{AppHPZ}

The HPZ equation has the Liouvillian
\begin{align}   \label{KHPZ}
    K_\HPZ(\w_0',b_\HPZ,d)&\equiv 2\w'_0 iL_0+\gam(O_0-I/2-iM_2)-2\gam b_\HPZ(O_++L_{1+})-dL_{2+}\,,
\end{align}
in the Markovian limit.
It is related by the following similarity transformation to the Liouvillian of the CL equation,
\begin{align}   \label{CLHPZ}
    K_\HPZ(\w'_0, b_\HPZ, d)
    &= e^{\zeta L_{1+}}K_\CL(\w'_0, b_\CL) e^{-\zeta L_{1+}}\,,
\end{align}
provided $b_\HPZ=b_\CL+\zeta/2$, and $d=-2\w'_0\zeta$.

Starting with the stationary state of the CL equation $f_{00}(Q,r)$ obtained in \ref{AppCL}, we apply item (6) from \ref{AppGauss} with $\mu=1/(4b_\CL), \mu+\nu=b_\CL, \kap=0$ to obtain
\begin{enumerate}[{(A)}]
\setcounter{enumi}{2}
\item $G_1(\zeta)f_{00}(Q,r)$.
\begin{subequations}
\begin{align}   \label{HPZA}
    \Del&=1\,,\\
    D&=1-\frac{\zeta}{2b_\CL}\,,\\
    \mu'&=\frac{1}{4(b_\CL-\zeta/2)}=\frac{1}{4\big[b_\HPZ+d/(2\w'_0)\big]}\,,\\
    \mu'+\nu'&=b_\CL+\frac{\zeta}{2}=b_\HPZ\,,\\
    \kap'&=0\,,\\
    b_\HPZ&=b_\CL-\frac{d}{4\w'_0}\,.
\end{align}
\end{subequations}
\end{enumerate}
In summary, the stationary state of $K_\HPZ(\w'_0,b_\HPZ,d)$ \eqref{KHPZ} is
\begin{align}   \label{f00HPZ}
    f'_{00}(Q,r)&=\frac{1}{\sqrt{2\pi(b_\HPZ+d/(2\w'_0))}}\,e^{-Q^2/[2(b_\HPZ+d/(2\w'_0))]-b_\HPZ r^2/2}\,.
\end{align}

\section{Examples of right eigenfunctions}
\label{Appright}

We begin by listing down the polynomial components of the first few right eigenfunctions \eqref{Pimn} of the KL equation. They are
\begin{subequations}
\begin{align}   \label{Pi11i}
    \tilde{\Pi}^\pm_{11}(\Qt,\rt)&= -i\big(\pm \Qt+\rt\big)\,,\\
    \tilde{\Pi}_{10}(\Qt,\rt)&= \frac{1}{2}-\Qt^2+\rt^2\,.
\end{align}
\end{subequations}
Notice that for $n=0, 2, 4, \cdots$, the $\pm$ superscripts on $\tilde{\Pi}^\pm_{mn}$ play no role. Hence, they are dropped.

The corresponding eigenvalues \eqref{lam} are
\begin{align}   \label{lambda11}
        \lambda^\pm_{11}&=\pm i\w_0+\gam/2\,,\qquad \lambda_{10}=\gam\,.
\end{align}

\subsection{Right eigenfunctions of Caldeira-Leggett (CL) equation}
\label{ApprightCL}

From \ref{AppCL}, we know that the CL equation can be similarly related to the KL equation by $S_1=U_1(\phi)G_2(\eta)$. A similarity transformation on the coordinates can be obtained by using the following equation,
\begin{align}   \label{SQrS}
    e^{A}Be^{-A}=B+[A,B]+\frac{1}{2!}[A,[A,B]]+\cdots\,.
\end{align}

\noindent (A) The commutator brackets between $L_{2+}$ and the position coordinates are
\begin{align}   \label{L2+Q}
    [L_{2+},Q]&=-\frac{i}{2}r\,,\qquad [L_{2+},r]=0\,,\qquad [L_{2+},\frac{\d}{\d r}]=\frac{i}{2}\frac{\d}{\d Q} \,,\qquad [L_{2+},\frac{\d}{\d Q}]=0\,.
\end{align}
Using \Eq{SQrS}, they yield
\begin{align}   \label{L2+QL2+}
    G_2(\eta)QG_2^{-1}(\eta)&=Q-\frac{i}{2} r\,, \qquad
    G_2(\eta)rG_2^{-1}(\eta)=r\,.
\end{align}

\noindent (B) In a similar way, the following commutator brackets between $iM_1$ and the position coordinates can be worked out
\begin{align}   \label{M1Q}
    [iM_1,Q]&=\frac{i}{2}\frac{\d}{\d r}\,, \qquad [iM_1,\frac{\d}{\d r}]=-\frac{i}{2}Q\,, \qquad [iM_1,r]= \frac{i}{2}\frac{\d}{\d Q}\,, \qquad     [iM_1,\frac{\d}{\d Q}]=-\frac{i}{2}r\,,
\end{align}
to give differential operators this time,
\begin{align}
\label{U1Q}
U_1(\phi)QU^{-1}(\phi)&=\cosh(\phi/2)Q+i\sinh(\phi/2)\frac{\d}{\d r}\,,\qquad
U_1(\phi)rU^{-1}(\phi)=\cosh(\phi/2)r+i\sinh(\phi/2)\frac{\d}{\d Q}\,.
\end{align}

Applying the results in (A) and (B) successively to the normalized coordinates $(\Qt,\rt)$ yield the differential operators
\begin{subequations}
\begin{align}   \label{UGQr}
    \hat{\Qb}&\equiv S_1\Qt S^{-1}_1 
            =\frac{1}{\sqrt{\cosh\phi}} \left(\cosh(\phi/2) \Qb
                    +\frac{i}{2}\sinh(\phi/2)\frac{\d}{\d \rb}\right)
                +\frac{\sinh\phi}{\sqrt{\cosh\phi}}\left( i\cosh(\phi/2) \rb
                     -\frac{1}{2} \sinh(\phi/2)\frac{\d}{\d \Qb}\right)\,,\\
    \hat{\rb}&\equiv S_1\rt S^{-1}_1 
            =\sqrt{\cosh\phi} \left(\cosh(\phi/2) \rb
                    +\frac{i}{2}\sinh(\phi/2)\frac{\d}{\d \Qb}\right)\,.
\end{align}
\end{subequations}
To facilitate the calculations, we have introduced the normalized coordinates for the CL equation
\begin{align}   \label{CLQr}
    \Qb\equiv \frac{Q}{\sqrt{2b_\CL}}\,,\qquad  \rb\equiv \sqrt{\frac{b_\CL}{2}} r\,.
\end{align}
The diifferential operators $\hat{\Qb}$ and $\hat{\rb}$ replace $\hat{Q}'$ and $\hat{r}'$ in \Eq{Sf}, respectively.
$\tilde{\Pi}^\pm_{mn}(\hat{Q}',\hat{r}')$ in \Eq{Sf} then becomes $\tilde{\Pi}^\pm_{mn}(\hat{\Qb},\hat{\rb})$ and acts on the stationary state
\begin{align}   \label{feqCL}
        f_\CL(\Qb,\rb)&=\frac{1}{\sqrt{2\pi b_\CL}}e^{-\Qb^2-\rb^2}\,.
\end{align}
to give the right eigenfunctions of the CL equation. $f_\CL(\Qb,\rb)$ is obtained in \ref{AppCL}.

The polynomial components of the first few eigenfunctions are
\begin{subequations}
\begin{align}   \label{f11CL}
        \bar{\Pi}^\pm_{11}(\Qb,\rb)&=\sqrt{\cosh\phi}\left(-\big[\sinh(\phi/2)\pm i\cosh(\phi/2)\big]\Qb+\big[\pm\sinh(\phi/2)-i\cosh(\phi/2)\big]\rb\right)\,,\\
        \bar{\Pi}_{10}(\Qb,\rb)&=\cosh\phi\left[\cosh\phi\left(\frac{1}{2}-\Qb^2+\rb^2\right)
                -2i\sinh\phi \,\Qb\rb\right]\,.
\end{align}
\end{subequations}
Using the definition of the hyperbolic functions in \ref{AppCL}, the polynomials in term sof the parameters of the master equation are
\begin{subequations}
\begin{align}   \label{f11CLw}
        \bar{\Pi}^+_{11}(\Qb,\rb)&=\frac{\sqrt{i\w_0'}}{\w_0} \left(i\sqrt{\lambda^-_{11}}\Qb
                                    +\sqrt{\lambda^+_{11}}\rb\right)\,,\\
        \bar{\Pi}^-_{11}(\Qb,\rb)&=\frac{\sqrt{i\w_0'}}{\w_0} \left(\sqrt{\lambda^+_{11}}\Qb
                                    -i\sqrt{\lambda^-_{11}}\rb\right)\,,\\
        \bar{\Pi}_{10}(\Qb,\rb)&=\frac{\w'_0}{\w_0} \left[\frac{\w'_0}{\w_0} \left(\frac{1}{2}-\Qb^2+\rb^2\right)
                +i\frac{\gam}{\w_0} \Qb\rb\right]\,,
\end{align}
\end{subequations}
in which $\lambda^\pm_{11}$ are the eigenvalues of the corresponding eigenfunctions \eqref{lambda11}.

The Liouvillian of the CL equation in the normalized position coordinates is
\begin{align}   \label{LCL}
        K_\CL(\Qb,\rb)=i\w_0'\left(-\frac{1}{2}\frac{\d^2}{\d\Qb\d\rb}+2\Qb\rb\right)
                    +\gam\rb\frac{\d}{\d \rb}+2\gam\rb^2\,.
\end{align}

\subsection{Right eigenfunctions of Hu-Paz-Zhang (HPZ) equation in the Markovian limit}
\label{ApprightHPZ}

The right eigenfunctions of the HPZ equation in the Markovian limit can be obtained from the right eigenfunctions of the CL equation through a similarity transformation initiated by $S_2=G_1(\zeta)$.

\noindent (C) We need the following commutator brackets
\begin{align}   \label{L1+Q}
    [L_{1+},Q]&=-\frac{1}{2}\frac{\d}{\d Q}\,,\qquad [L_{1+},\frac{\d}{\d Q}]=0\,,
    \qquad[L_{1+},r]=0\,, \qquad [L_{1+},\frac{\d}{\d r}]=\frac{1}{2}r  \,.
\end{align}
to obtain
\begin{align}   \label{L1+QL1+}
    S_2 Q S_2^{-1}=Q-\frac{\zeta}{2}\frac{\d}{\d Q}\,, \qquad
    S_2 r S_2^{-1}=r\,.
\end{align}
Introducing the normalized position coordinates for HPZ equation,
\begin{align}   \label{HPZQr}
    Q_+&\equiv \frac{Q}{\sqrt{2b_+}}\,,\qquad\qquad\qquad\qquad\qquad  r_-\equiv \sqrt{\frac{b_-}{2}} r\,,\\
    b_+&\equiv b_\HPZ+\frac{d}{2\w_0'} =b_\CL+ \frac{d}{4\w_0'}\,, \qquad
    b_-\equiv b_\HPZ=b_\CL- \frac{d}{4\w_0'}\,.
\end{align}
They produce the transformed differential operators
\begin{subequations}
\begin{align}   \label{Qhat+}
    \hat{Q}_+&\equiv S_2\Qb S_2^{-1}
            =\frac{1}{\sqrt{2b_+(b_++b_-)}}
            \left[2b_+Q_++\frac{1}{2}(b_+-b_-)\frac{\d}{\d Q_+}\right] \,, \\
    \hat{r}_-&\equiv S_2\rb S_2^{-1}=\sqrt{\frac{b_++b_-}{2b_-}}r_-\,,
\end{align}
\end{subequations}
We substitute $\hat{Q}_+, \hat{r}_-$ for $\Qb,\rb$, respectively, in the polynomial components $\Pi\bar{\Pi}^\pm_{mn}(\Qb,\rb)$ of the eigenfunctions of the CL equation.
$\bar{\Pi}^\pm_{mn}(\hat{Q}_+, \hat{r}_-)$ then acts on the stationary state of the HPZ equation
\begin{align}   \label{feqHPZ}
        f_\HPZ(Q_+,r_-)&=\frac{1}{\sqrt{2\pi b_+}}e^{-Q_+^2-r_-^2}\,,
\end{align}
to give the right eigenfunctions of the HPZ equation. The polynomial components are, for example,
\begin{subequations}
\begin{align}   \label{f11hpzw}
        \Pi^+_{11}(Q_+,r_-)&=\frac{\sqrt{i\w_0'}}{\w_0}\sqrt{b_++b_-}
                        \left(i\sqrt{\frac{\lambda^-_{11}}{2b_+}}Q_+
                            +\sqrt{\frac{\lambda^+_{11}}{2b_-}}r_-\right)\,,\\
        \Pi^-_{11}(Q_+,r_-)&=\frac{\sqrt{i\w_0'}}{\w_0}\sqrt{b_++b_-} \left(\sqrt{\frac{\lambda^+_{11}}{2b_+}}Q_+
                    -i\sqrt{\frac{\lambda^-_{11}}{2b_-}}r_-\right)\,,\\
        \Pi_{10}(Q_+,r_-)&=\frac{\w'_i0}{\w_0}\frac{b_++b_-}{2b_+}
                \left[\frac{\w'_0}{\w_0} \left(\frac{1}{2}-Q_+^2+\frac{b_+}{b_-} r_-^2\right)
                +i\frac{\gam}{\w_0}\sqrt{\frac{b_+}{b_-}}Q_+r_-\right]\,.
\end{align}
\end{subequations}

The Liouvillian of the HPZ equation in the normalized position coordinates is
\begin{align}   \label{LHPZ}
        K_\HPZ(Q_+,r_-)=i\w_0'\left(-\frac{1}{2}\sqrt{\frac{b_-}{b_+}}\frac{\d^2}{\d Q_+\d r_-}
        +2\sqrt{\frac{b_+}{b_-}} Q_+r_-\right)
                    +\gam\rb\frac{\d}{\d \rb}+2\gam\rb^2+i\w'_0\frac{b_+-b_-}{\sqrt{b_+b_-}}r_-\frac{\d}{\d Q_+}\,.
\end{align}

\providecommand{\noopsort}[1]{}\providecommand{\singleletter}[1]{#1}%

\end{document}